\documentclass[10pt,journal,compsoc, onecolumn]{IEEEtran}

\usepackage[nocompress]{cite}
\usepackage{graphicx}
\usepackage{xcolor}
\usepackage{array}

\begin{document}

\title{How far are we from quantifying visual attention in mobile  HCI?}

\author{Mihai~B\^ace, ETH Z\"urich, 
        Sander~Staal, ETH Z\"urich, 
        Andreas~Bulling, University of Stuttgart 
\IEEEcompsocitemizethanks{
\IEEEcompsocthanksitem M. B\^ace is with the Department
of Computer Science, ETH Z\"urich, Switzerland.\protect\\
E-mail: mihai.bace@inf.ethz.ch
\IEEEcompsocthanksitem S. Staal is with the Department
of Computer Science, ETH Z\"urich, Switzerland.
\IEEEcompsocthanksitem A. Bulling is with the Institute for Visualisation and Interactive Systems, University of Stuttgart, Germany.
}}

\IEEEtitleabstractindextext{
%!TEX root = ../paper.tex
\begin{abstract}
    With an ever-increasing number of mobile devices competing for our attention,
    quantifying when, how often, or for how long users visually attend to their devices has emerged as a core challenge in mobile human-computer interaction.
    Encouraged by recent advances in automatic eye contact detection using machine learning and device-integrated cameras, we provide a fundamental investigation into the feasibility of quantifying visual attention during everyday mobile interactions.
    We identify core challenges and sources of errors associated with sensing attention on mobile devices in the wild, including the impact of face and eye visibility, the importance of robust head pose estimation, and the need for accurate gaze estimation.
    Based on this analysis, we propose future research directions and discuss how eye contact detection represents the foundation for exciting new applications towards next-generation pervasive attentive user interfaces.
\end{abstract}

}

\maketitle

\IEEEdisplaynontitleabstractindextext

\IEEEpeerreviewmaketitle

%!TEX root = ../paper.tex

\section{Introduction}
\label{sec:introduction}

In recent years, the number of digital interfaces competing for users' limited and, as such, valuable attentional resources has rapidly increased. 
Consequently, actively managing user attention has emerged as a fundamental research challenge in human-computer interaction (HCI).
With mobile devices being pervasively used in daily life, this challenge is even more relevant in mobile HCI where attentive behaviour has, as a result, become highly fragmented~\cite{Oulasvirta:2005:IBF:1054972.1055101,Steil:2018:FUA:3229434.3229439}.
A first step and key requirement to better understand and actively manage attention is to quantify when, how often, or for how long users visually attend to their devices. 

Only few previous works have tried to quantify user attention and its allocation on mobile devices despite its fundamental importance for tasks such as predicting interruptibility~\cite{Pejovic:2014:IDI:2632048.2632062} or modeling and understanding smartphone application usage~\cite{Jones:2015:RAS:2750858.2807542}.
Most of these studies required special-purpose eye tracking equipment~\cite{Steil:2018:FUA:3229434.3229439} that constrained users' mobility or manual data annotation~\cite{Oulasvirta:2005:IBF:1054972.1055101} that prevented the study of natural attentive behaviour at scale.
One way to overcome these limitations is to instead rely on the high-resolution front-facing cameras readily integrated into these devices in combination with computer vision for gaze estimation~\cite{zhang15_cvpr, cvpr2016_gazecapture, zhang2017s, zhang18_pami}.
However, despite significant progress in recent years,
such methods are still too inaccurate to analyse attention in a fine-grained manner, e.g. in terms of individual eye movements. 

Eye contact has been proposed as an alternative given that, although being coarser and easier to detect, it still provides rich insights into attentive behaviour.
In contrast to gaze estimation where the goal is to predict a precise 3D gaze direction or 2D location on a screen, eye contact detection is
the binary task of detecting if users look at their devices or not. 
Encouraged by recent advances in automatic eye contact detection~\cite{Zhang:2017:EEC:3126594.3126614}, for the first time, we study the feasibility of quantifying visual attention during everyday interactions with mobile devices.
More specifically, to guide future research on this emerging topic, we identify key challenges and the most important sources of error.
We first evaluate the impact of face and eye visibility on eye contact detection performance, given that the best performing methods require face and facial landmarks but the users' face is only visible around 30\% of the time by device-integrated cameras in mobile settings~\cite{Khamis:2018:UFE:3173574.3173854}.
We then study the impact of head pose on eye contact detection performance, which is particularly challenging in mobile settings in which devices are held and being looked at in a variety of ways, including while on the go.
Finally, we discuss the need for more accurate gaze estimation and its importance to the eye contact detection task. 
For each of these challenges, we propose concrete future research directions and show how eye contact detection can form the basis for higher-level attention metrics that will enable a range of exciting new applications towards pervasive attentive user interfaces~\cite{bulling16_computer}. 
%!TEX root = ../paper.tex

\section{Attention Analysis}\label{sec:related-work}

Over the years, several approaches have been proposed to sense overt attention, also typically referred to as just visual attention.
These can be categorized into two groups: Methods that rely on special-purpose hardware as well as software-only methods that only require commodity devices.

Early work in the first group are EyePliances~\cite{eyepliances}, which are custom-built devices that respond to human attention. 
Equipped with a camera, they detect eye contact using computer vision techniques by detecting a pupil in the image.
A similar idea has been proposed for human-to-human eye contact detection in the form of eye contact sensing glasses~\cite{Dickie:2004:ECS:985921.985927} that are equipped with infrared cameras and LEDs.
Recently, commercial eye trackers have become smaller and widely accessible which makes them good candidates for attention analysis~\cite{Kassner:2014:POS:2638728.2641695}. 
Steil et al. have used such an eye tracker together with mobile device-integrated sensors to forecast user attention~\cite{Steil:2018:FUA:3229434.3229439}. 
While such dedicated systems bring us closer to the vision of pervasive attentive user interfaces, the fact that they require special-purpose equipment hinders large-scale deployment.

In contrast, software-only methods leverage the fact that mobile devices have ever-increasing computational capabilities and readily integrated, sophisticated sensors. 
As such, these methods do not require any custom hardware as the camera is already available in these devices, and they enable studying attention in-situ, i.e. during users' everyday interactions. 
Integrated cameras, in particular, have improved significantly in recent years in terms of resolution and quality and now enable visual computing methods for attention analysis unthinkable before. 
For example, EyeTab was an early approach to estimate users' gaze direction during interactions with a tablet device~\cite{Wood:2014:EMG:2578153.2578185}.
Their system required only the front-facing RGB camera and achieved an angular error of around 6$^{\circ}$.
Learning-based gaze estimation methods are more promising because they can learn parameters from large datasets to create person-independent gaze estimators.
One such approach is the full-face appearance-based gaze estimator proposed by Zhang et al.\cite{zhang2017s} that uses a Convolutional Neural Network (CNN) trained on the MPIIGaze dataset~\cite{zhang15_cvpr}.
A similar approach proposed specifically for mobile devices is iTracker~\cite{cvpr2016_gazecapture}.
While such appearance-based gaze estimation methods have improved significantly and can achieve gaze estimation errors of around 4-6$^{\circ}$, these methods are still less accurate than dedicated eye trackers.

Hence, a third line of work investigated eye contact detection as a computationally simpler variation of the gaze estimation task that focuses on coarser events in users' attentive behaviour, such as when, how often, or for how long users look at their devices.
Gaze locking is a fully supervised approach for appearance-based eye contact detection~\cite{Smith:2013:GLP:2501988.2501994}, however, manually creating and annotating such datasets is tedious and impractical. 
To address this limitation, Zhang et al.~\cite{Zhang:2017:EEC:3126594.3126614} proposed an alternative method for eye contact detection that, besides achieving state-of-the-art performance, is unsupervised, i.e. does not require manual annotation.
%!TEX root = ../paper.tex

\section{Everyday Eye Contact Detection}
\label{sec:method}

The single assumption of the method by Zhang et al. is that the camera is next to the object of interest -- an assumption that is also true for common mobile devices. 
The first step involves detecting the user's face with a CNN face detector, included in the dlib library.
Afterwards, a landmark detector finds six landmarks inside the face bounding box: the four corners of the eyes and the two mouth corners.
These six 2D facial landmarks together with the corresponding 3D points, taken from a generic 3D facial shape model~\cite{zhang15_cvpr}, can be used to estimate the 3D head pose by solving a Perspective-n-Point (PnP) problem.
Then, the input image is normalized by rotating and scaling it to a space with fixed camera parameters, which is beneficial to the gaze estimation task~\cite{zhang18_etra}.
After image normalization, a state-of-the-art gaze estimation CNN~\cite{zhang2017s} is used to predict the gaze direction vector.
Intersecting this vector with the camera plane will produce the corresponding 2D gaze location.
These 2D gaze locations are sampled for clustering under the assumption that each cluster corresponds to one eye contact target. 
Since the camera is always placed next to the object of interest, the correct data cluster is the one closest to the camera, i.e. closest to the origin of the coordinate system. 

After clustering, samples belonging to the target cluster will be labeled as positive, while  all the others will be labeled as negative.
These images can now be used to train a binary Support Vector Machine (SVM) as the eye contact classifier. 
The SVM input is a 4096-dimensional feature vector extracted from the last fully-connected layer of the gaze estimation CNN. 
Clustering is only necessary once, for training. 
For inference, input images are still pre-processed and fed into the same gaze estimation CNN model. 
The trained SVM classifier then takes the feature vector as input and outputs the predicted eye contact label.
%!TEX root = ../paper.tex

\section{Key Challenges in Quantifying Mobile Visual Attention}\label{sec:challenges}

The purpose of our work is to provide a fundamental analysis of using the approach by Zhang et al. for quantifying visual attention during everyday mobile interactions.
To this end, in our implementation of their method, we used the dlib\footnote{http://dlib.net} CNN face detector, the dlib 68 landmark detector, and we trained a full-face appearance-based gaze estimator on the MPIIFaceGaze dataset~\cite{zhang2017s}. 
We conducted our experiments on two challenging and publicly available datasets:
\begin{itemize}
	\item \textit{Understanding Face and Eye Visibility Dataset (UFEV)}~\cite{Khamis:2018:UFE:3173574.3173854}. 
	It consists of 25,726 images collected by 10 participants during everyday in-the-wild mobile interactions. 
	The objective of the dataset was to analyze the visibility of the different facial landmarks, such as eyes or mouth, when users naturally interact with their mobile device. 
	For our evaluation, we sampled 5,791 images which were manually annotated with eye contact labels. 
	4844 images were labeled as positive eye contact labels and the remaining 947 images were labeled with negative no eye contact labels. 
	
	\medskip
	
	\item \textit{Mobile Face Video Dataset (MFV)}~\cite{mfv-dataset}.
	It aims to provide a better understanding of the challenges associated with mobile face-based authentication. 
	While a different computational task than gaze estimation or eye contact detection, this dataset is nevertheless interesting because it contains 750 face videos from 50 users in different illumination conditions captured using the front-facing camera of an iPhone 5s. 
	While collecting the data, users had to perform five different tasks.
	In our experiments, we select the "enrollment" task where users had to turn their heads in four different directions (up, down, left, and right).
	This enabled us to create a more balanced evaluation dataset (as opposed to the UFEV dataset).
	Out of the 4,363 manually annotated images, around 58\% were positive eye contact labels and the remaining were negative labels.
	
\end{itemize}

Before investigating the different factors that influence the accuracy and robustness of the method, we first evaluated the overall performance in terms of the Matthews Correlation Coefficient (MCC), which is commonly used to asses binary classifiers.
The MCC ranges from -1.0, which indicates total contradiction between the predictions and the observations, to 1.0, which corresponds to a perfect classifier.
A value of 0 is equivalent to random guessing.
Overall, on the UFEV dataset, the method achieves an MCC of 0.349 (SD=0.17) in a leave-one-person-out cross validation. 
We also evaluated the performance of the method in an ablation study where we assumed perfect eye contact labels.
In such a case, the method's MCC score increases to 0.499 (SD=0.17).

Within-dataset evaluations only highlight one aspect of performance. 
With machine learning systems, it is also interesting to asses them across datasets, which is a good indicator of real-world performance. 
In this experiment, we trained the eye contact detector on one dataset and evaluated its performance on the other.
Training on MFV and evaluating on UFEV, the MCC score is 0.124.
Assuming perfect eye contact labels, the MCC score increases to 0.403.
Training on UFEV and testing on MFV, the MCC score is 0.484.
With ground truth labels, the MCC score is 0.431. 

To better understand the failure cases, we then identified and studied three core challenges: Partially visible faces, the impact of different head pose angles, and gaze estimation performance as a basis for eye contact detection.

\begin{figure}[!ht]
	\centering
	\includegraphics[width=\textwidth]{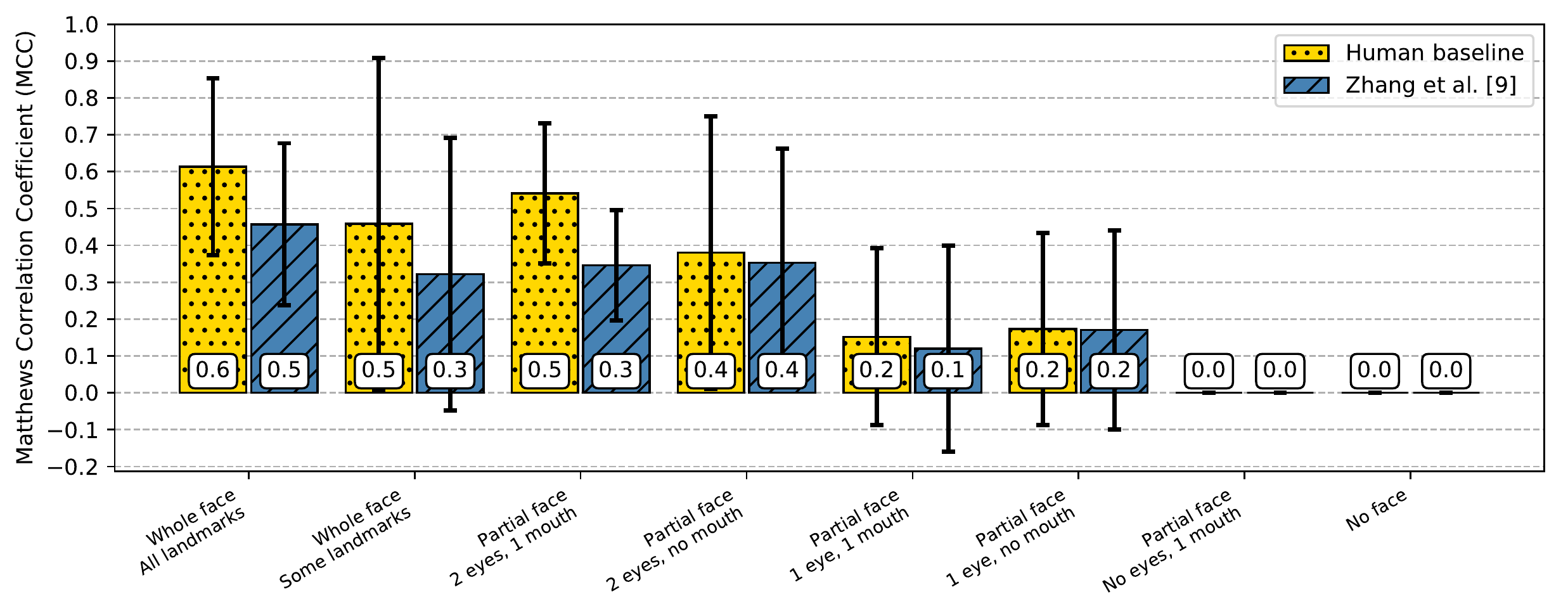}
	\caption{Performance of the two methods, the eye contact detector by Zhang et al. and the Human baseline which assumes ground truth class labels. The bars represent the MCC and the error bars represent the standard deviation. The results are from a within dataset evaluation on the UFEV dataset (leave-one-person-out per visibility category cross validation).}
	\label{fig:mcc-within-percategory}
\end{figure}

\subsection{Challenge 1: Face and Eye (In)visibility}\label{sec:evaluation}

One highly relevant challenge for studies conducted using the front-facing camera of mobile devices is the face and eye visibility of the participants~\cite{Khamis:2018:UFE:3173574.3173854}. 
Nowadays, most face detection, landmark detection, and even many gaze estimation approaches require the full face to be visible.
However, according to Khamis et al.~\cite{Khamis:2018:UFE:3173574.3173854}, the full face is only visible around 30\% of the time. 
The method by Zhang et al.~\cite{Zhang:2017:EEC:3126594.3126614} also requires the full face as input given that one of the steps in their pipeline is a full-face appearance-based gaze estimator. 
This is why, in this section, we evaluate the impact of partially visible faces on the method's performance. 

Our evaluation is conducted on the UFEV dataset which provides several different visibility categories depending on whether the entire face or only parts of the face are visible. 
These categories are: \textit{Whole face all landmarks}, \textit{Whole face some landmarks}, \textit{Partial face 2 eyes 1 mouth}, \textit{Partial face 2 eyes no mouth}, \textit{Partial face 1 eye 1 mouth}, \textit{Partial face 1 eye no mouth}, \textit{Partial face no eyes 1 mouth}, and \textit{No face}.

Figure~\ref{fig:mcc-within-percategory} shows the result of a within dataset leave-one-person-out per category cross validation.
The rightmost two categories, \textit{Partial face no eyes 1 mouth} and \textit{No face}, have an MCC score of 0 simply because no images could be used in the evaluation, either because no faces were detected or because all the images only belonged to a single class. 
Thus, it is not possible to train and evaluate a classifier.
For the remaining categories, we compare the method proposed by Zhang et al.~\cite{Zhang:2017:EEC:3126594.3126614} to the same method when using ground truth labels, the Human baseline. 
The results are as follows.
When the full face is visible, the MCC is 0.457 (SD=0.22). 
In the Human baseline, the MCC is 0.613 (SD=0.24), which shows the potential for improving the unsupervised clustering approach for automatic labeling of the data. 
For the other categories, the MCC score degrades when fewer landmarks are visible. 
If two eyes are visible, the average MCC stays above 0.3, however, once only one eye or less is visible, the method simply becomes unusable. 

\begin{figure}[!ht]
	\centering
	\includegraphics[width=\textwidth]{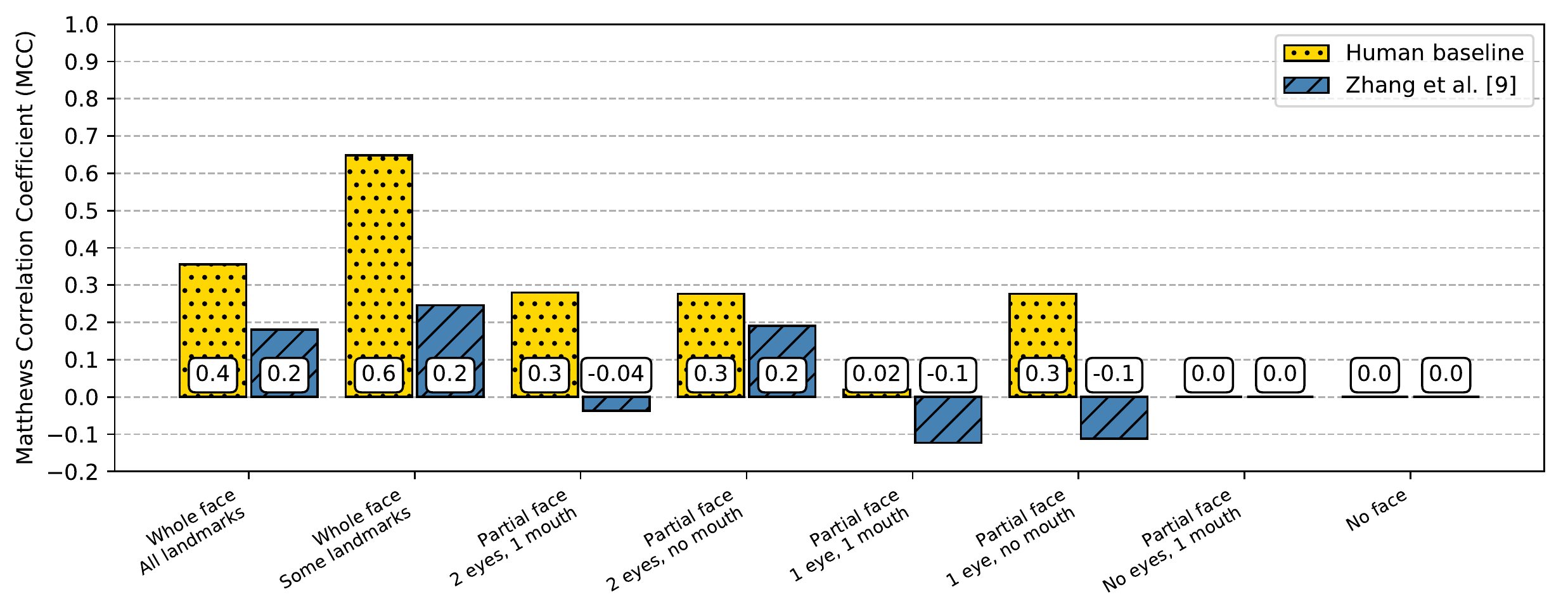}
	\caption{Performance of the two methods, the eye contact detector by Zhang et al. and the Human baseline which assumes ground truth class labels. The bars represent the MCC coefficient. The results are from a cross-dataset evaluation where the eye contact detector was trained on the entire MFV dataset and tested on the UFEV dataset for all participants, per category.}
	\label{fig:mcc-crossdataset-percategory}
\end{figure}

To understand real-world performance, we conducted a cross-dataset evaluation (see
Figure~\ref{fig:mcc-crossdataset-percategory} for a performance overview of the method). 
The eye contact detector was trained on the MFV dataset and evaluated once on the entire UFEV dataset, per visibility category.
In this case, it becomes even clearer that the method performs poorly and could be significantly improved when comparing its performance with the human baseline.

\subsection{Challenge 2: Robust Head Pose Estimation}

\begin{figure}[!t]
	\centering
	\includegraphics[width=.9\textwidth]{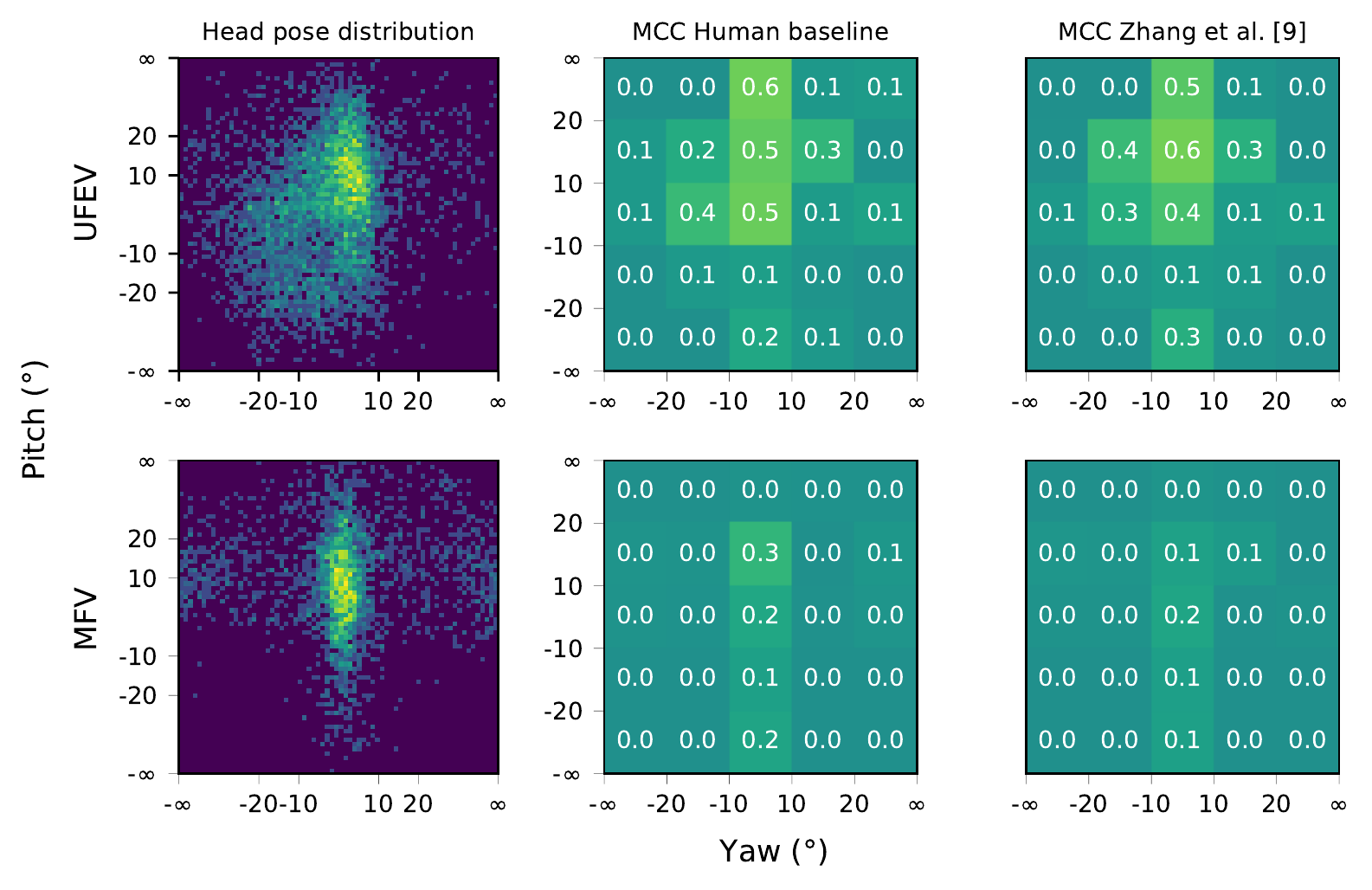}
	\caption{Classification performance of the eye contact detector by head pose angles. The left most column shows the distribution of the pitch and yaw in the normalized camera space. The MCC values represent the performance of the two baselines, per bucket, from a leave-one-person-out cross validation. The Human baseline uses ground truth annotations rather than clustering to obtain the labels for the training samples.}
	\label{fig:mcc-headpose}
\end{figure}

Head pose estimation is a computer vision task where the goal is to determine how the head is tilted relative to the camera.
It is expressed in terms of six degrees of freedom, three for translation and three for rotation in 3D. 
For the appearance-based gaze estimation task, head pose estimation is often used as input to train a CNN or for data normalization~\cite{zhang18_etra}. 
In mobile settings (see Figure~\ref{fig:mcc-headpose} - Head pose distribution), for both datasets, we have noticed a large variability in both the horizontal and the vertical pose angles.
Because of this, we investigated the influence of such angles on the eye contact detection performance.
In other words, is the performance of eye contact detection worse when the head is tilted and not frontal? Does this happen often in mobile scenarios? 

Figure~\ref{fig:mcc-headpose} shows the results of this experiment. 
The first column represents the distribution of the head pose angles in the normalized camera space~\cite{zhang18_etra} estimated from the two datasets. 
For the experiments, we divided the data in five horizontal and five vertical buckets. 
A pitch and yaw value between -10$^{\circ}$ and 10$^{\circ}$ represent little rotation of the head.
Between 10$^{\circ}$ and 20$^{\circ}$ is a mild turn of the head.
We consider anything over 20$^{\circ}$ as a significant head rotation.
As shown in the head pose distribution, in mobile settings, it is often the case that the head and face are not directly facing the camera.

The reported values represent the MCC coefficient from a within dataset leave-one-person-out per bucket cross validation.
The first row highlights the result on the UFEV dataset, while the second one shows the results on the MFV dataset. 
On the UFEV dataset, for pitch and yaw values between -10$^{\circ}$ and 10$^{\circ}$, the MCC score is 0.4 for Zhang et al. and 0.5 when using ground truth labels. 
Because of the distribution of the data, a similar MCC value is achieved when the pitch is between 10$^{\circ}$ and 20$^{\circ}$. 
As the angles become more extreme, the methods become unusable. 
On the MFV dataset, the performance is even worse. 
For frontal faces, the MCC value for Zhang et al. is 0.2.

\subsection{Challenge 3: Accurate Gaze Estimation}

\begin{figure}[h]
	\centering
	\includegraphics[width=\textwidth]{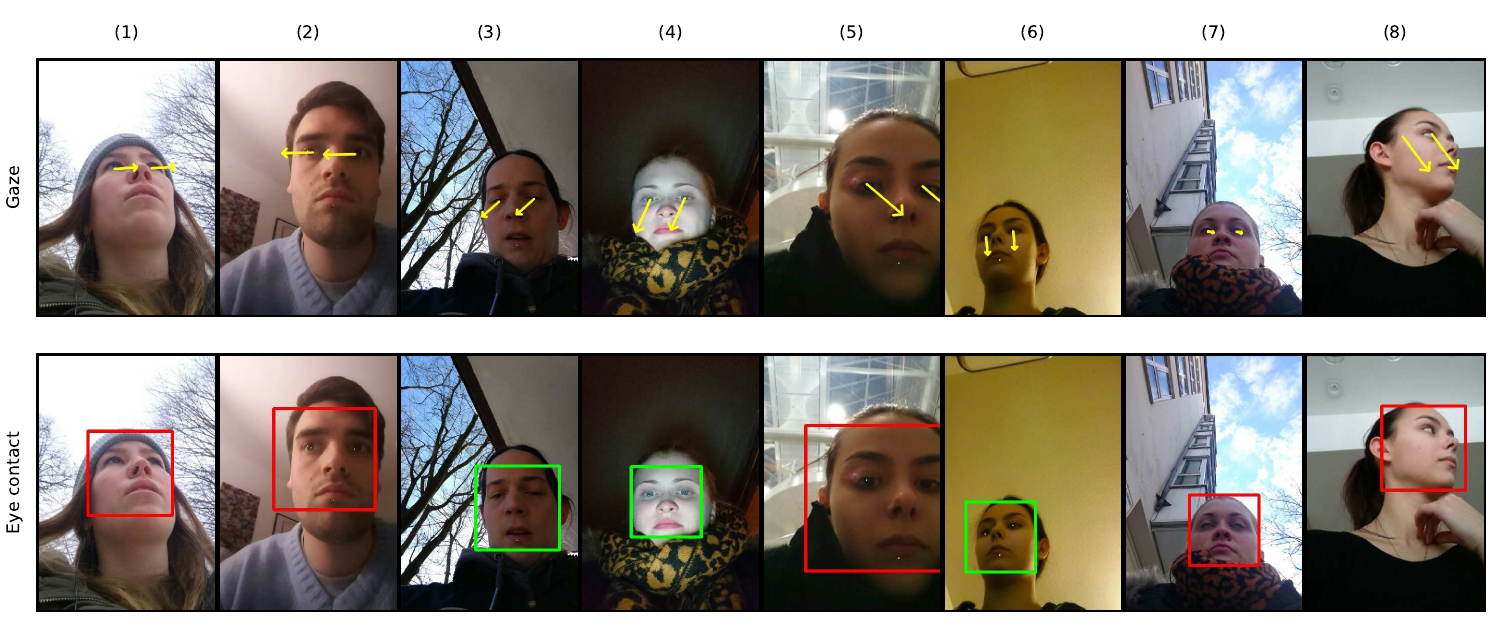}
	\caption{Sample images with the corresponding gaze estimates and the predicted eye contact label (green represents eye contact, red non eye contact). While being computationally simpler, the state-of-the-art method proposed by Zhang et al. for eye contact detection builds on an appearance-based gaze estimator. Thus, the performance of the method is dependent on the performance of the underlying gaze estimates. E.g. for certain head poses (column 6), if the gaze estimates are incorrect, the eye contact label will also be incorrect.}
	\label{fig:image-samples}
\end{figure}

Recent advances in appearance-based gaze estimation bring us closer to the vision of systems that are able to accurately track human gaze from a single image~\cite{Zhang:2017:EEC:3126594.3126614, zhang15_cvpr, zhang18_pami, cvpr2016_gazecapture}. 
Despite these advancements, most gaze estimators are still far from practical use due to lower accuracies and the eye contact detection method proposed by Zhang et al. builds on such an appearance-based gaze estimator.
Consequently, improvements to the gaze estimation task will also benefit eye contact detection. 
Estimating the gaze direction in everyday settings has to cope with several challenges.
Varying illumination conditions, variability across users, different screen and camera geometries, face and facial landmarks occlusions are only a few of the challenges which have to be addressed for accurate and robust gaze estimation. 
Figure~\ref{fig:image-samples} shows a few sample images from the UFEV dataset together with the gaze estimates and the predicted eye contact label. 
For some images, Figure~\ref{fig:image-samples} columns 1-4, if the gaze estimates are reasonably accurate, the method is able to overcome small estimation errors and correctly predict (no) eye contact.
However, gaze estimates can also be highly inaccurate if, for example, the face and facial landmarks have been incorrectly detected (column 8).
Another possible source of error is due to the head pose angles (column 6).
Most current gaze estimation datasets only contain limited variability in head pose angles, but as seen in Figure~\ref{fig:mcc-headpose}, mobile settings can exhibit a wide range of head orientations.
Without additional training data, the predicted gaze estimates in such cases will be inaccurate as well.

%!TEX root = ../paper.tex

\section{Discussion}\label{sec:discussion}

In our evaluations, we identified three key challenges for sensing attention in highly dynamic, mobile interactive settings.

Our first experiment quantified the impact of face and eye visibility on the eye contact classification performance and showed that current methods performed best when the full face or all the facial landmarks were visible.
As soon as the eyes or parts of them, which convey most of the relevant information for attention, were not visible, the performance of the method decreased significantly and became unusable. 
This is also a consequence of methods which rely on the full face for training appearance-based gaze estimators.
While suitable for desktop settings, just as the findings from Khamis et al.~\cite{Khamis:2018:UFE:3173574.3173854} highlight, in mobile settings the entire face is often not visible. 
As such, future work should continue to investigate methods that only require, for example, an image of the eye~\cite{zhang18_pami} rather than the entire face, or that are at least able to work robustly with a smaller number of facial landmarks.
As such, another critical research direction is to investigate more robust algorithms for face detection and landmark localization. 
In our evaluation, the face detector failed around 30\% of the time for images with partial faces even though both eyes were visible.
In more extreme cases where at most one eye was visible, the failure rate was as high as 81\%.

Our second experiment on the error distribution of the eye contact detector relative to the distribution of the head pose angles yielded several interesting findings.
For one, current methods perform best when the head is oriented towards the camera. 
As soon as the head is turned in any direction, the performance of the method becomes worse.
However, we can observe that if there is sufficient training data available for such cases, e.g. Figure~\ref{fig:mcc-headpose} - on the UFEV dataset when the pitch is larger than $10^\circ$, the method can still perform well. 
Based on this, as future research directions, we believe that at least two things are important.
First, the head pose angles we used are estimates (there is no ground truth available), so it is possible that some of these are incorrect or inaccurate.
Future research could investigate head pose estimation in mobile settings and asses accuracy and robustness specifically.
Second, there is a need for new datasets that cover a variety of not only head pose angles but gaze angles as well. 

Our last experiment qualitatively addressed the need for accurate gaze estimation. 
As previously mentioned, eye contact detection methods, while computationally simpler, still require reasonable gaze estimates to produce usable results. 
As such, any improvement in current gaze estimation methods will also benefit attention sensing on mobile devices. 
More concretely, we encourage future work to investigate gaze estimation methods and datasets which have been collected specifically in such mobile interactive scenarios (e.g. the large-scale GazeCapture dataset~\cite{cvpr2016_gazecapture}). 

\medskip 

Our analysis, so far, shows that there is still a large gap that has to be filled before attention can be sensed accurately and robustly in mobile settings. 
Once some of these challenges have been addressed, we envision several application domains that can benefit from knowing when, how often, or for how long users attend to their devices. 
On the one hand, eye contact detection can be used as a means to sense and quantify attentive behavior during everyday mobile interactions. 
Just as in the work by Steil et al.~\cite{Steil:2018:FUA:3229434.3229439}, eye contact could be used as a basis for higher-level attention metrics. 
Such metrics could count the number of times users attend to their device, for how long, or if they have shifted their attention towards the environment. 
These, together with other device-integrated sensors, would enable modeling user behaviour in a way which is currently not possible without special-purpose eye trackers. 
These user models could also be used for other tasks in mobile HCI, such as predicting user interruptibility, assessing user engagement or boredom.
On the other hand, real-time eye contact detection could be used for attentive and interactive user interfaces. 
For instance, if users do not look at their device, the screen could be turned off to save power. 
Some mobile device manufacturers already offer a similar functionality, however, this is so far only based on head pose information and, as such, error-prone. 
Another possible application is in the area of quantified self. 
Both Apple and Android smartphones quantify the amount of time users spend on their devices.
Such statistics are naively based on the amount of time the screen is on, however, with eye contact detection, much finer insights could be provided. 
For example, attentive behaviour and the way users interact while using social media could be completely different than while browsing the Internet or while texting.
%!TEX root = ../paper.tex

\section{Conclusion}\label{sec:conclusion}

In this work, we investigated the feasibility of quantifying visual attention during everyday mobile interactions. 
To this end, for the first time, we studied a state-of-the-art method for automatic eye contact detection in challenging mobile interactive scenarios. 
We identified three core challenges associated with sensing attention in the wild and provided future research directions for each of them: Face and eye (in)visibility, robust head pose estimation, and the need for accurate gaze estimation. 
Last but not least, we discussed how eye contact (detection) and attention quantification on mobile devices will enable exciting new applications.
As such, our work informs the development of future pervasive attentive user interfaces and provides concrete guidance for researchers and practitioners working in this emerging research area alike.

\bibliographystyle{IEEEtran}
\bibliography{references}

\begin{thebibliography}{10}
\providecommand{\url}[1]{#1}
\csname url@samestyle\endcsname
\providecommand{\newblock}{\relax}
\providecommand{\bibinfo}[2]{#2}
\providecommand{\BIBentrySTDinterwordspacing}{\spaceskip=0pt\relax}
\providecommand{\BIBentryALTinterwordstretchfactor}{4}
\providecommand{\BIBentryALTinterwordspacing}{\spaceskip=\fontdimen2\font plus
\BIBentryALTinterwordstretchfactor\fontdimen3\font minus
  \fontdimen4\font\relax}
\providecommand{\BIBforeignlanguage}[2]{{%
\expandafter\ifx\csname l@#1\endcsname\relax
\typeout{** WARNING: IEEEtran.bst: No hyphenation pattern has been}%
\typeout{** loaded for the language `#1'. Using the pattern for}%
\typeout{** the default language instead.}%
\else
\language=\csname l@#1\endcsname
\fi
#2}}
\providecommand{\BIBdecl}{\relax}
\BIBdecl

\bibitem{Oulasvirta:2005:IBF:1054972.1055101}
A.~Oulasvirta, S.~Tamminen, V.~Roto, and J.~Kuorelahti, ``{Interaction in
  4-second Bursts: The Fragmented Nature of Attentional Resources in Mobile
  HCI},'' in \emph{Proceedings of the SIGCHI Conference on Human Factors in
  Computing Systems}, ser. CHI '05.\hskip 1em plus 0.5em minus 0.4em\relax New
  York, NY, USA: ACM, 2005, pp. 919--928,
  http://doi.acm.org/10.1145/1054972.1055101.

\bibitem{Steil:2018:FUA:3229434.3229439}
J.~Steil, P.~M\"{u}ller, Y.~Sugano, and A.~Bulling, ``Forecasting user
  attention during everyday mobile interactions using device-integrated and
  wearable sensors,'' in \emph{Proceedings of the 20th International Conference
  on Human-Computer Interaction with Mobile Devices and Services}, ser.
  MobileHCI '18.\hskip 1em plus 0.5em minus 0.4em\relax New York, NY, USA: ACM,
  2018, pp. 1:1--1:13, http://doi.acm.org/10.1145/3229434.3229439.

\bibitem{Pejovic:2014:IDI:2632048.2632062}
V.~Pejovic and M.~Musolesi, ``Interruptme: Designing intelligent prompting
  mechanisms for pervasive applications,'' in \emph{Proceedings of the 2014 ACM
  International Joint Conference on Pervasive and Ubiquitous Computing}, ser.
  UbiComp '14.\hskip 1em plus 0.5em minus 0.4em\relax New York, NY, USA: ACM,
  2014, pp. 897--908, http://doi.acm.org/10.1145/2632048.2632062.

\bibitem{Jones:2015:RAS:2750858.2807542}
S.~L. Jones, D.~Ferreira, S.~Hosio, J.~Goncalves, and V.~Kostakos,
  ``Revisitation analysis of smartphone app use,'' in \emph{Proceedings of the
  2015 ACM International Joint Conference on Pervasive and Ubiquitous
  Computing}, ser. UbiComp '15.\hskip 1em plus 0.5em minus 0.4em\relax New
  York, NY, USA: ACM, 2015, pp. 1197--1208,
  http://doi.acm.org/10.1145/2750858.2807542.

\bibitem{zhang15_cvpr}
X.~Zhang, Y.~Sugano, M.~Fritz, and A.~Bulling, ``Appearance-based gaze
  estimation in the wild,'' in \emph{Proc. of the IEEE Conference on Computer
  Vision and Pattern Recognition (CVPR 2015)}, 2015, pp. 4511--4520.

\bibitem{cvpr2016_gazecapture}
K.~Krafka, A.~Khosla, P.~Kellnhofer, H.~Kannan, S.~Bhandarkar, W.~Matusik, and
  A.~Torralba, ``Eye tracking for everyone,'' in \emph{IEEE Conference on
  Computer Vision and Pattern Recognition (CVPR)}, 2016.

\bibitem{zhang2017s}
X.~Zhang, Y.~Sugano, M.~Fritz, and A.~Bulling, ``It’s written all over your
  face: Full-face appearance-based gaze estimation,'' in \emph{Computer Vision
  and Pattern Recognition Workshops (CVPRW), 2017 IEEE Conference on}.\hskip
  1em plus 0.5em minus 0.4em\relax IEEE, 2017, pp. 2299--2308.

\bibitem{zhang18_pami}
------, ``Mpiigaze: Real-world dataset and deep appearance-based gaze
  estimation,'' \emph{IEEE Transactions on Pattern Analysis and Machine
  Intelligence (TPAMI)}, vol.~41, no.~1, pp. 162--175, 2019.

\bibitem{Zhang:2017:EEC:3126594.3126614}
X.~Zhang, Y.~Sugano, and A.~Bulling, ``Everyday eye contact detection using
  unsupervised gaze target discovery,'' in \emph{Proceedings of the 30th Annual
  ACM Symposium on User Interface Software and Technology}, ser. UIST
  '17.\hskip 1em plus 0.5em minus 0.4em\relax New York, NY, USA: ACM, 2017, pp.
  193--203, http://doi.acm.org/10.1145/3126594.3126614.

\bibitem{Khamis:2018:UFE:3173574.3173854}
M.~Khamis, A.~Baier, N.~Henze, F.~Alt, and A.~Bulling, ``Understanding face and
  eye visibility in front-facing cameras of smartphones used in the wild,'' in
  \emph{Proceedings of the 2018 CHI Conference on Human Factors in Computing
  Systems}, ser. CHI '18.\hskip 1em plus 0.5em minus 0.4em\relax New York, NY,
  USA: ACM, 2018, pp. 280:1--280:12,
  http://doi.acm.org/10.1145/3173574.3173854.

\bibitem{bulling16_computer}
A.~Bulling, ``Pervasive attentive user interfaces,'' \emph{IEEE Computer},
  vol.~49, no.~1, pp. 94--98, 2016.

\bibitem{eyepliances}
J.~S. Shell, R.~Vertegaal, and A.~W. Skaburskis, ``Eyepliances:
  Attention-seeking devices that respond to visual attention,'' in \emph{CHI
  '03 Extended Abstracts on Human Factors in Computing Systems}, ser. CHI EA
  '03.\hskip 1em plus 0.5em minus 0.4em\relax New York, NY, USA: ACM, 2003, pp.
  770--771, http://doi.acm.org/10.1145/765891.765981.

\bibitem{Dickie:2004:ECS:985921.985927}
C.~Dickie, R.~Vertegaal, J.~S. Shell, C.~Sohn, D.~Cheng, and O.~Aoudeh, ``Eye
  contact sensing glasses for attention-sensitive wearable video blogging,'' in
  \emph{CHI '04 Extended Abstracts on Human Factors in Computing Systems}, ser.
  CHI EA '04.\hskip 1em plus 0.5em minus 0.4em\relax New York, NY, USA: ACM,
  2004, pp. 769--770, http://doi.acm.org/10.1145/985921.985927.

\bibitem{Kassner:2014:POS:2638728.2641695}
M.~Kassner, W.~Patera, and A.~Bulling, ``Pupil: An open source platform for
  pervasive eye tracking and mobile gaze-based interaction,'' in
  \emph{Proceedings of the 2014 ACM International Joint Conference on Pervasive
  and Ubiquitous Computing: Adjunct Publication}, ser. UbiComp '14
  Adjunct.\hskip 1em plus 0.5em minus 0.4em\relax New York, NY, USA: ACM, 2014,
  pp. 1151--1160, http://doi.acm.org/10.1145/2638728.2641695.

\bibitem{Wood:2014:EMG:2578153.2578185}
E.~Wood and A.~Bulling, ``Eyetab: Model-based gaze estimation on unmodified
  tablet computers,'' in \emph{Proceedings of the Symposium on Eye Tracking
  Research and Applications}, ser. ETRA '14.\hskip 1em plus 0.5em minus
  0.4em\relax New York, NY, USA: ACM, 2014, pp. 207--210,
  http://doi.acm.org/10.1145/2578153.2578185.

\bibitem{Smith:2013:GLP:2501988.2501994}
B.~A. Smith, Q.~Yin, S.~K. Feiner, and S.~K. Nayar, ``Gaze locking: Passive eye
  contact detection for human-object interaction,'' in \emph{Proceedings of the
  26th Annual ACM Symposium on User Interface Software and Technology}, ser.
  UIST '13.\hskip 1em plus 0.5em minus 0.4em\relax New York, NY, USA: ACM,
  2013, pp. 271--280, http://doi.acm.org/10.1145/2501988.2501994.

\bibitem{zhang18_etra}
X.~Zhang, Y.~Sugano, and A.~Bulling, ``Revisiting data normalization for
  appearance-based gaze estimation,'' in \emph{Proc. International Symposium on
  Eye Tracking Research and Applications (ETRA)}, 2018, pp. 12:1--12:9.

\bibitem{mfv-dataset}
M.~E. {Fathy}, V.~M. {Patel}, and R.~{Chellappa}, ``Face-based active
  authentication on mobile devices,'' in \emph{2015 IEEE International
  Conference on Acoustics, Speech and Signal Processing (ICASSP)}, April 2015,
  pp. 1687--1691.

\end{thebibliography}

\begin{IEEEbiographynophoto}{Mihai B\^ace}
is a PhD student and research assistant at the Institute for Pervasive Computing, Department of Computer Science, ETH Z\"urich, Switzerland. His research interests are at the intersection of machine learning, computer vision, and human-computer interaction with a focus on eye tracking.
\end{IEEEbiographynophoto}

\begin{IEEEbiographynophoto}{Sander Staal}
is a graduate MSc student and research assistant at the Institute for Pervasive Computing, Department of Computer Science, ETH Z\"urich, Switzerland. His main research interests include ubiquitous computing, computer vision, and human-computer interaction.
\end{IEEEbiographynophoto}

\begin{IEEEbiographynophoto}{Andreas Bulling}
is Full Professor of Computer Science at the University of Stuttgart where he holds the chair for Human-Computer Interaction and Cognitive Systems. He received his MSc. in Computer Science from the Karlsruhe Institute of Technology, Germany, in 2006 and his PhD in Information Technology and Electrical Engineering from ETH Zurich, Switzerland, in 2010. He was previously a Feodor-Lynen and Marie Curie Research Fellow at the University of Cambridge, UK, and a Senior Researcher at the Max Planck Institute for Informatics, Germany. His research interests include computer vision, wearable and ubiquitous computing, and human-computer interaction.
\end{IEEEbiographynophoto}

\end{document}